\documentclass{article}
\usepackage{spconf,amsmath,amssymb,graphicx}
\usepackage{booktabs}
\usepackage{multirow}
\usepackage{array}
\usepackage{xcolor}
\usepackage{colortbl}
\usepackage{makecell}
\usepackage{tikz}
\usetikzlibrary{positioning,fit,backgrounds,arrows.meta,calc}
\usepackage{balance}

\title{Indic-CLAP: Text--Audio Understanding in Indic Languages \\
       via Cross-Lingual Distillation}
\name{Sanat Kumar Agrawal and Srikanth Raj Chetupalli}
\address{Electrical Engineering, Indian Institute of Technology Bombay, Mumbai, India \\
         \texttt{\{sanat.agrawal, srikanthrajch\}@iitb.ac.in}
         }

\begin{document}
\ninept
\maketitle

\begin{abstract}
Contrastive Language--Audio Pretraining (CLAP) learns a joint text--audio
representation space, which enables zero-shot audio understanding from
natural-language descriptions. However, its English-only text encoder restricts CLAP to tasks and datasets expressed in English. We present
Indic-CLAP, a multilingual text--audio model that extends CLAP to nine Indic
languages spoken by over a billion people. Exploiting CLAP's dual-encoder
structure, we propose to train only an Indic text encoder by distillation while keeping the audio encoder frozen, using machine-translated AudioCaps captions. We compare distillation from CLAP's text encoder, audio encoder, and a novel hybrid objective combining both. Our evaluation shows that the Indic-CLAP encoder inherits the cross-modal alignment from the CLAP teacher, as evidenced by the performance on cross-modal retrieval and zero-shot classification tasks. Further, we study the cross-modal alignment using modality gap analysis, which explains the performance trends.

\end{abstract}


\section{Introduction}
Contrastive Language--Audio Pretraining (CLAP) \cite{elizalde2023clap} maps audio and text into a joint representation space learned from audio recordings paired with natural-language descriptions. The resulting model is shown to have zero-shot capabilities, with applications in audio classification and tagging
\cite{elizalde2023clap, wu2023largescaleclap}, captioning
\cite{deshmukh2023pengi}, text--audio retrieval \cite{koepke2022retrieval}, and
text-guided audio generation \cite{liu2023audioldm, ghosal2023tango}. Further, the model was extended to yield richer audio and speech representations in \cite{wu2023largescaleclap,dinkel2025glap,sun25c_interspeech} to  support tasks like keyword spotting and emotion recognition. 

\noindent CLAP is inspired by the text-image representation model CLIP \cite{radford2021clip} and uses a dual-encoder structure with its text branch built on RoBERTa \cite{liu2019roberta}, a monolingual (English) representation model. Further, CLAP is trained only on audio--caption pairs written in English \cite{wu2023largescaleclap}, limiting its usage. GLAP \cite{dinkel2025glap} added multilingual capabilities by training on translated captions, but in only four non-English languages, none of them are from the Indic languages. Other approaches in literature, such as RA-CLAP~\cite{sun2025raclap}, also do not have multilingual capabilities. The translated-caption approach was first
studied in \cite{yan24_interspeech} for text--audio retrieval and in
\cite{cousin2023multilingualaudiocaptioningusing} for audio captioning.
 However, similar works for Indic languages are non-existent in the literature. To the best of our knowledge, there are no public audio
datasets with captions in Indic languages, and no existing model is trained or evaluated for Indic-language audio--text tasks. Indic NLP has grown rapidly through multilingual representations \cite{khanuja2021muril, kakwani2020indicnlpsuite}, but audio--language tasks remain largely unexplored. We try to bridge this gap in this work, by developing CLAP-based models for nine Indic languages. 


\noindent Our approach is inspired by M-CLIP \cite{carlsson2022mclip} and extends the linguistic capabilities of CLAP by exploiting its dual-encoder structure and training a multilingual text encoder to mimic the default English encoder. We apply this {\it cross-lingual distillation} approach to Indic languages and propose Indic-CLAP, an extension of CLAP to nine languages. The multilingual text encoder is trained on translated captions via distillation from the original text and audio encoders of CLAP. The audio encoder is frozen and the distillation based training ensures the semantic alignment with Indic-CLAP text embeddings by transitivity. 


\noindent Our contributions are as follows. (i) We propose Indic-CLAP, which covers Hindi, Bengali, Tamil, Telugu, Marathi, Gujarati, Kannada, Malayalam and Punjabi. We study language-specific models and a joint multilingual model, and show that joint training substantially improves downstream task performacen over language-specific models. (ii) We compare a neural machine translation (NMT) pipeline against an LLM-based pipeline and show that the performance is sensitive to the translation quality.
(iii) We study four distillation objectives, which includes a hybrid objective that distills from both CLAP encoders. Our results show that the hybrid objective gives the best cross-modal retrieval, which is explained by a mechanistic account based on the cross-modal misalignment (modality gap).
(iv) We extend the approach to GLAP \cite{dinkel2025glap} and show that the distillation strategy is independent of the architecture. 

\section{Methodology}
\label{sec:method}

\subsection{Cross-Lingual Teacher Learning}
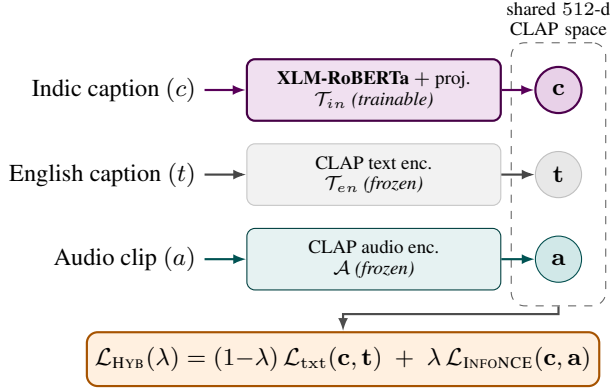
\begin{figure}[t]
\centering
\resizebox{\columnwidth}{!}{%
\begin{tikzpicture}[
  every node/.style={font=\small},
  inp/.style={anchor=east, minimum height=5mm, font=\small},
  enc/.style={draw, rounded corners=3pt, minimum width=33mm,
    minimum height=8mm, align=center, font=\scriptsize},
  encs/.style={enc, fill=violet!12, draw=violet!55!black, thick},
  enct/.style={enc, fill=gray!10,    draw=gray!45},
  enca/.style={enc, fill=teal!10,    draw=teal!50!black},
  emb/.style={draw, circle, minimum size=6mm, inner sep=0pt, font=\small\bfseries},
  embs/.style={emb, fill=violet!18, draw=violet!60!black, thick},
  embt/.style={emb, fill=gray!18,   draw=gray!50},
  emba/.style={emb, fill=teal!18,   draw=teal!60!black},
  loss/.style={draw, rounded corners=4pt, minimum height=7mm, align=center,
    fill=orange!12, draw=orange!65!black, thick, font=\small},
  ar/.style={-{Latex[length=1.7mm,width=1.3mm]}, line width=0.8pt},
]
\node[inp] (indic) at (12mm,   0mm) {Indic caption $(c)$};
\node[inp] (encap) at (12mm, -11mm) {English caption $(t)$};
\node[inp] (audio) at (12mm, -22mm) {Audio clip $(a)$};

\node[encs] (xlmr) at (34mm,   0mm) {\textbf{XLM-RoBERTa} $+$ proj.\\$\mathcal{T}_{in}$ \emph{(trainable)}};
\node[enct] (ften) at (34mm, -11mm) {CLAP text enc.\\$\mathcal{T}_{en}$ \emph{(frozen)}};
\node[enca] (faen) at (34mm, -22mm) {CLAP audio enc.\\$\mathcal{A}$ \emph{(frozen)}};

\node[embs] (s) at (58mm,   0mm) {$\mathbf{c}$};
\node[embt] (t) at (58mm, -11mm) {$\mathbf{t}$};
\node[emba] (a) at (58mm, -22mm) {$\mathbf{a}$};

\begin{scope}[on background layer]
  \node[draw=black!60, dashed, rounded corners=5pt, inner xsep=3mm, inner ysep=3mm,
        fit=(s)(t)(a)] (embbox) {};
\end{scope}
\node[font=\scriptsize, align=center, text width=22mm] at (58mm, 9mm)
      {shared $512$-d\\CLAP space};

\draw[ar, violet!70!black] (indic) -- (xlmr);
\draw[ar, violet!70!black] (xlmr)  -- (s);
\draw[ar, gray!55!black]   (encap) -- (ften);
\draw[ar, gray!55!black]   (ften)  -- (t);
\draw[ar, teal!65!black]   (audio) -- (faen);
\draw[ar, teal!65!black]   (faen)  -- (a);

\node[loss] (lhyb) at (30mm, -35mm)
  {$\mathcal{L}_{\textsc{Hyb}}(\lambda)=(1{-}\lambda)\,\mathcal{L}_{\mathrm{txt}}(\mathbf{c},\mathbf{t})
     \;+\;\lambda\,\mathcal{L}_{\textsc{InfoNCE}}(\mathbf{c},\mathbf{a})$};
\draw[ar, black!70] (embbox.south) -- ++(0,-1mm) -| (lhyb.north);
\end{tikzpicture}}

\caption{Cross-lingual CLAP distillation. Only the Indic student
$\mathcal{T}_{in}$ is trained; both CLAP encoders stay frozen and their
embeddings are pre-computed, so no audio is read at training time. 
}
\label{fig:arch}

\end{figure}

CLAP~\cite{elizalde2023clap} follows a dual-encoder architecture in which the audio encoder
($\mathcal{A}$) and the English text encoder ($\mathcal{T}_{en}$) are
independent and interact only through a shared embedding space ($\mathcal{E}$)
on the unit sphere. $\mathcal{E}$ is learned by aligning the encoder outputs over many audio--text description pairs using a contrastive loss~\cite{radford2021clip}. Our main idea is to train a text encoder
($\mathcal{T}_{in}$) for Indic languages that generates embeddings in
$\mathcal{E}$, inheriting alignment with the outputs of $\mathcal{A}$, as shown in Figure~\ref{fig:arch}.
Let $\{(a, t, c)\}$ be a triplet of audio, English caption and
its Indic-language caption, and let $\mathbf{a} = \mathcal{A}(a)$ and
$\mathbf{t} = \mathcal{T}_{en}(t)$ be the CLAP embeddings, with
$\mathbf{a}, \mathbf{t} \in \mathcal{E}$. The goal is to learn
$\mathcal{T}_{in}$ such that its output $\mathbf{c} =
\mathcal{T}_{in}(c)$ is in $\mathcal{E}$ and the pairwise cosine
similarity (CS) between the elements of the triplet $(\mathbf{a}, \mathbf{t}, \mathbf{c})$ is high.

\noindent Existing audio-captioning datasets contain only English captions, so we resort to machine translation. That is, given $t$, the Indic caption is obtained as
$\hat{c} = \mathrm{translate}(t)$, and the resulting triplets $({a}, {t}, \hat{c})$ are used to train $\mathcal{T}_{in}$ in a teacher--student framework with $\mathcal{A}$ and
$\mathcal{T}_{en}$ as teachers. Since only $\mathcal{T}_{in}$ is trained, the pairs $\{(\mathbf{a}, \mathbf{t})\}$ are pre-computed and cached for all examples in the dataset.

\subsection{Indic Text Encoder ($\mathcal{T}_{in}$)}
The architecture of $\mathcal{T}_{in}$ closely follows CLAP's text encoder $\mathcal{T}_{en}$. We
use pre-trained XLM-RoBERTa-base \cite{conneau2020xlmr} with attention-masked
mean pooling over token states followed by a linear projection and $\ell_2$
normalization as the text encoder. We chose XLM-RoBERTa-base, as it was pre-trained on $100$ languages, which include the nine Indic languages considered in this work. Training only the linear projection was found to be insufficient in our preliminary experiments, hence the model was finetuned end-to-end after loading the pretrained weights of XLM-RoBERTa-base\footnote{\scriptsize \texttt{hf.co/FacebookAI/xlm-roberta-base}}.


\subsection{Training Objectives}
\label{sec:objectives}
We explore four distillation objectives, summarized in Figure~\ref{fig:arch}. The two mean squared error (MSE) objectives, text-teacher MSE (TT-MSE) and audio-teacher MSE (AT-MSE), use the CLAP's text and audio encoders as teachers, respectively. Hence, in the TT-MSE case, the Indic-CLAP student $\mathcal{T}_{in}$ is trained to mimic $\mathcal{T}_{en}$, 
\begin{equation}
  \mathcal{L}_{\textsc{TT-MSE}} (\mathbf{t}, \hat{c}) = 
  \|\mathbf{t} - \mathcal{T}_{in}(\hat{c})\|_2^2 ,
\end{equation}
and alignment with the audio encoder $\mathcal{A}$ is expected by transitivity.
In the AT-MSE case, $\mathcal{T}_{in}$ is
trained to mimic the audio encoder,
\begin{equation}
  \mathcal{L}_{\textsc{AT-MSE}} (\mathbf{a},\hat{c}) = 
  \|\mathbf{a} - \mathcal{T}_{in}(\hat{c})\|_2^2 .
\end{equation}
This is a harder problem, since it distills across modalities. As shown in
Section~\ref{sec:modality_gap_analysis}, CLAP suffers from modality
misalignment and AT-MSE yields a poor text encoder.

\noindent Symmetric InfoNCE is the contrastive objective used to train CLAP \cite{radford2021clip, oord2018cpc}. Hence, we use the same objective, the symmetric cross-entropy loss over the similarity matrix, to train the Indic-CLAP text encoders against the frozen audio encoder. We refer to this as the audio-teacher InfoNCE (AT-NCE) objective.
Unlike AT-MSE, AT-NCE aligns embeddings semantically. However, significant misalignment may remain between the encoder pairs. In particular, the trained Indic-CLAP encoder $\mathcal{T}_{in}$ need not be aligned with $\mathcal{T}_{en}$.

\noindent To address the modality alignment with both the text and audio encoders of CLAP, 
we propose distillation from both the encoders in a hybrid objective,
\begin{multline}
\label{eq:hybrid}
  \mathcal{L}_{\textsc{Hyb},\lambda}(\{\mathbf{t}_i\},\{\hat{c}_i\}) =
  (1{-}\lambda)\frac{1}{B}\sum_{i=1}^B
  \bigl(1 - \cos(\mathbf{t}_i, \mathcal{T}_{in}(\hat{c}_i))\bigr) \\
  + \lambda\,\mathcal{L}_{\textsc{InfoNCE}}
  (\{\mathbf{a}_i\}, \{\mathcal{T}_{in}(\hat{c}_i)\}),
\end{multline}
where the two terms measure semantic alignment with CLAP's text and audio encoders, $\lambda \in [0,1]$ is a hyperparameter, and the metric is defined for a batch of $B$ samples. Cosine similarity is used for the text term because it is scale-commensurate with InfoNCE, unlike the Euclidean distance.

\begin{table*}[t]
\centering
\small
\setlength{\tabcolsep}{8pt}
\caption{Language-specific vs.\ multilingual Indic-CLAP, both trained with the
TT-MSE objective. In matched settings the CLAP teacher achieves R@1 of $24.5\%$
(A$\rightarrow$T) and $24.2\%$ (T$\rightarrow$A), and ZSC accuracy of $92.5\%$.}
\label{tab:tt_perlang_vs_joint}
\vspace{2pt}
\begin{tabular}{l cccc cccc}
\toprule
 & \multicolumn{4}{c}{\textbf{Language-specific Indic-CLAP}}
 & \multicolumn{4}{c}{\textbf{Multilingual Indic-CLAP}} \\
\cmidrule(lr){2-5}\cmidrule(lr){6-9}
\textbf{Lang.} & DF & A$\rightarrow$T & T$\rightarrow$A & ZSC
               & DF & A$\rightarrow$T & T$\rightarrow$A & ZSC \\
 & (R@1) & (R@1) & (R@1) & (Acc.) & (R@1) & (R@1) & (R@1) & (Acc.) \\
\midrule
Hindi (hi)     & 63.6 & 33.0 & 21.6 & 84.8 & 74.7 & 29.8 & 23.2 & 81.9 \\
Bengali (bn)   & 61.5 & 30.8 & 20.9 & 83.0 & 73.2 & 30.3 & 22.7 & 80.5 \\
Marathi (mr)   & 58.5 & 31.1 & 20.6 & 79.0 & 71.4 & 30.4 & 22.3 & 79.5 \\
Gujarati (gu)  & 61.0 & 30.0 & 21.0 & 74.8 & 73.2 & 29.5 & 23.1 & 70.0 \\
Punjabi (pa)   & 57.2 & 29.9 & 21.0 & 78.3 & 71.3 & 29.6 & 22.4 & 74.5 \\
Tamil (ta)     & 50.6 & 29.2 & 18.9 & 67.0 & 65.1 & 28.8 & 20.8 & 74.8 \\
Telugu (te)    & 57.9 & 31.2 & 20.4 & 71.7 & 70.5 & 29.7 & 22.1 & 74.9 \\
Kannada (kn)   & 51.9 & 28.9 & 19.4 & 70.2 & 66.4 & 27.6 & 21.0 & 73.4 \\
Malayalam (ml) & 54.4 & 31.6 & 19.2 & 71.2 & 68.3 & 29.8 & 21.4 & 68.2 \\
\midrule
\textbf{Avg.} & \textbf{57.4} & \textbf{30.6} & \textbf{20.3} & \textbf{75.6}
              & \textbf{70.5} & \textbf{29.5} & \textbf{22.1} & \textbf{75.3} \\
\bottomrule
\end{tabular}
\vspace{-10pt}
\end{table*}

\vspace{-5pt}
\section{Experimental Setup}
\vspace{-5pt}

We used the CLAP model pretrained on an extended caption set
\cite{wu2023largescaleclap}, which combines LAION-Audio-630K with AudioCaps
\cite{kim2019audiocaps} and Clotho
\cite{drossos2020clotho}, as the teacher for training. The model pairs an
HTSAT \cite{chen2022htsat} audio encoder with a RoBERTa \cite{liu2019roberta}
text encoder, and was contrastively trained with InfoNCE to produce a shared $512$-dimensional space.
For Indic-CLAP, we trained both, the language-specific (a separate model for each language), and the multilingual models. All models were trained using AdamW, with a linear warm-up learning rate scheduling for $1$K steps, peak learning rate of $2{\times}10^{-5}$, and cosine annealing to $10^{-7}$. The batch size used was $256$ and the gradients were clipped to a global $\ell_2$ norm of $1.0$. Early stopping was used with a maximum of $30$~epochs. For the MSE objectives, early stopping was set to 
monitor validation loss with patience $5$, while the InfoNCE-based objectives (AT-NCE and the hybrid) monitored validation retrieval performance (R@1) with patience $5$ and $8$, respectively. 
In all cases, the linear layer weights were initialized from $\mathcal{N}(0, 0.0004)$ with zero bias. The teacher CLAP embeddings from the audio and text encoders were pre-computed, so no audio was processed at training time.


\noindent Our experiments are based on the AudioCaps \cite{kim2019audiocaps} dataset, which contains $52.25$K AudioSet \cite{gemmeke2017audioset} recordings with manually written English captions, split into $\{45.72\textrm{K}, 2.17\textrm{K}, 4.36\textrm{K}\}$ samples for train, val, and test, respectively. AudioCaps is CLAP's primary training set, so the teacher embeddings are well
calibrated. Training uses translated captions, which is difficult because audio captions are terse and sound-specific. Hence, we studied caption translation using Helsinki-NMT \cite{tiedemann2020opus}\footnote{\scriptsize
\texttt{hf.co/Helsinki-NLP/opus-mt-en-hi}} and OpenAI \texttt{gpt-4o-mini}
\cite{openai2024gpt4}. 




\noindent We evaluate the downstream performance using cross-modal retrieval and zero-shot classification tasks, as well as the distillation fidelity (DF). DF quantifies alignment between the student and the CLAP text teacher. In this, each translated test caption is first encoded by the student, then ranked by CS against all teacher embeddings of the test set, and the top-1 recall (R@1 in \%) is reported.
The \emph{cross-modal retrieval} experiments evaluated the performance on both, text-to-audio (T$\rightarrow$A), and audio-to-text (A$\rightarrow$T) retrieval tasks. Queries (text/audio) were first encoded, then matched by CS against the CLAP embeddings (audio/text), and the top-1 (R@1) and top-10 (R@10) retrieval accuracies are reported. We performed the retrieval experiments on two datasets, (i) AudioCaps test set containing $970$ unique recordings with $4{,}360$ captions, and (ii) the Clotho test set with $1{,}045$ unique recordings. 

\noindent The \emph{Zero-shot classification} (ZSC) experiments were conducted on the ESC-50~\cite{piczak2015esc} test set containing $2$K recordings from $50$ classes and the UrbanSound8K dataset having $375$ recordings from $10$ classes. Note that the test sets used for ZSC are not seen by the model during training. For ZSC, the translated class labels are embedded in natural-language prompt templates (e.g.\ the Hindi template \emph{``yah \{label\} ki aavaz hai''} meaning, ``This is a sound of \{label\}'') and the CLAP embeddings were first computed for each class label. Two native Hindi speakers independently audited the translated ESC-50 and UrbanSound8K class labels; $59$ of the $60$ were judged correct, with one label (\emph{thunderstorm}) marked different, but not wrong. During evaluation, each clip is first passed through the CLAP audio encoder and then compared against the class label embeddings using CS as the metric to assign the class. Top-1 accuracy (\%) is then reported as the performance metric.

\noindent We compared the Indic-CLAP models with GLAP \cite{dinkel2025glap} in a zero-shot setting, where GLAP is directly tested on the nine Indic languages, and after finetuning GLAP's text encoder, using the objectives described in Section~\ref{sec:objectives}. GLAP \cite{dinkel2025glap} was neither trained nor evaluated on any Indic language, but its SONAR \cite{duquenne2023sonar} text branch covers all nine languages considered in this work, allowing a direct comparison under our protocol in zero-shot setting. SONAR was trained with a machine translation objective to learn language-agnostic sentence level representations, and we hypothesize that this enables GLAP to perform in the languages it was not trained in a zero-shot manner. 

\vspace{-5pt}
\section{Results}
\vspace{-5pt}
\label{sec:results}

\subsection{Translation Quality}

\label{sec:translation}

\begin{table}[t]
\centering
\small
\setlength{\tabcolsep}{7pt}
\caption{Effect of caption translation quality evaluated for the Hindi Indic-CLAP encoder trained with the TT-MSE objectives. 
}
\label{tab:translation_quality}
\vspace{2pt}
\begin{tabular}{l cccc}
\toprule
\textbf{Translation} & \textbf{DF} & \textbf{A$\rightarrow$T} & \textbf{T$\rightarrow$A}
                     & \textbf{ZSC} \\
 & (R@1) & (R@1) & (R@1) & (Acc.) \\
\midrule
Helsinki-NMT & 37.2 & 25.5 & 14.7 & 57.7 \\
\rowcolor{blue!8}
GPT-4o-mini & \textbf{63.6} & \textbf{33.0} & \textbf{21.6} & \textbf{84.8} \\
\bottomrule
\end{tabular}
\vspace{-10pt}
\end{table}

We assessed the translation quality by training the Hindi Indic-CLAP encoder with the TT-MSE objective on the translations of AudioCaps captions by two different translation systems. Table~\ref{tab:translation_quality} shows the performance measures, which shows that the choice of translator is critical.
The model trained on Helsinki-NMT translated captions has poor DF compared to GPT-4o-mini and also a poor ZSC accuracy.
The same ordering also holds for both the retrieval tasks.
We conclude that the translation quality is a key design choice and limits the performance of student on every task we measure. 


\noindent Audio captions are linguistically unusual. They are short, onomatopoeic (for example \emph{crackling}, \emph{chirping}, \emph{sizzling}) and rich in compound descriptors. Helsinki-NMT is trained on general parallel corpora and we observed that, it often leaves sound words untranslated, maps domain terms to unrelated meanings (\emph{sizzling} $\rightarrow$ ``visit happens''), misses sound-event ordering, and drops audio-specific vocabulary. 
GPT-4o-mini, prompted with audio context, preserves onomatopoeia and sound-event structure, leading to better alignment with the CLAP teacher. 
Hence, we use GPT-4o-mini for caption translation in all languages. The text prompt used for the caption translation, along with translated captions and demonstrative examples are available online.\footnote{ https://github.com/Sanat-agrwl/IndicCLAP-csvs}\footnote{ https://sanat-agrwl.github.io/IndicCLAP} 

\vspace{-3pt}
\subsection{Multilingual vs.\ Language-Specific Models}
\vspace{-2pt}
Table~\ref{tab:tt_perlang_vs_joint} compares Indic-CLAP encoders trained separately per language against a single jointly trained multilingual encoder. TT-MSE is used as the training objective for the models studied in this section. DF is substantially better for the multilingual encoder, reaching upto $75\%$ retrieval accuracy, showing that text encoders for new languages can be aligned with CLAP's English encoder and that the joint training for multiple languages is superior to language specific encoders. The retrieval tasks show an interesting trend; on the A$\rightarrow$T task, the Indic-CLAP encoders \emph{exceed} the CLAP teacher performance and the performance is uncorrelated with DF, whereas the T$\rightarrow$A task performance correlates with DF showing a degradation relative to the teacher. We attribute this to the asymmetry in the datasets used for training, where several text captions map to a single audio sample. The A$\rightarrow$T task is easy, since the audio embedding needs to be similar to one of the text captions, compared to the T$\rightarrow$A task where the text embedding from Indic-CLAP needs to match exactly that of English-CLAP for a successful retrieval. ZSC accuracy is similar for both encoder types, with a gap of $\approx 17$ points to the English CLAP model. We attribute this to the gap in  DF; which saturates near $75\%$, so the student does not fully reproduce the teacher's concept structure leading to a reduced zero-shot classification performance.


\subsection{Training Objectives}
\begin{table}[t]
\centering
\small
\setlength{\tabcolsep}{3.5pt}
\caption{Performance of multilingual Indic-CLAP averaged over all nine languages, compared with CLAP, and variants of GLAP finetuned to Indic languages.
}
\label{tab:objectives_lambda}
\vspace{3pt}
\begin{tabular}{l c cc cc}
\toprule
 & & \multicolumn{2}{c}{\textbf{A$\rightarrow$T (\%)}}
     & \multicolumn{2}{c}{\textbf{T$\rightarrow$A (\%)}} \\
\cmidrule(lr){3-4}\cmidrule(lr){5-6}
\textbf{Objective} & \textbf{ZSC (\%)} & \textbf{R@1} & \textbf{R@10}
                   & \textbf{R@1} & \textbf{R@10} \\
\midrule
TT-MSE & 75.3 & 29.5 & 74.7 & 22.1 & 71.9 \\
AT-MSE & 60.8 & 27.3 & 77.2 & 17.9 & 63.9 \\
AT-NCE & 54.7 & 28.2 & 78.3 & 23.4 & 69.9 \\
\midrule
\multicolumn{6}{l}{\textit{Hybrid objective, sweep over $\lambda$}} \\
0.00 & \textbf{75.6} & 29.0 & 74.7 & 22.0 & 71.9 \\
\rowcolor{blue!8}
\textbf{0.10} & 73.3 & \textbf{38.8} & \textbf{84.9} & \textbf{25.4} & \textbf{75.4} \\
0.25 & 68.6 & 36.3 & 83.2 & 25.1 & 74.4 \\
0.50 & 65.9 & 32.3 & 81.1 & 24.9 & 72.8 \\
0.75 & 56.5 & 29.9 & 79.6 & 24.7 & 71.7 \\
0.90 & 55.9 & 30.0 & 78.8 & 23.3 & 70.5 \\
1.00 & 54.7 & 28.2 & 78.3 & 23.4 & 69.9 \\
\midrule
\rowcolor{gray!12}
CLAP (teacher) & 92.5 & 24.5 & 74.9 & 24.2 & 74.7 \\
\rowcolor{gray!12}
GLAP \cite{dinkel2025glap} (English) & 87.8 & 52.5 & 90.7 & 40.1 & 81.7 \\
\rowcolor{gray!12}
GLAP \cite{dinkel2025glap} (zero shot) & 38.5 & 40.6 & 85.0 & 28.2 & 69.4 \\
\quad $+$ TT-MSE & 75.0 & 50.3 & 89.6 & 36.1 & 79.2 \\
\quad $+$ Hybrid $\lambda{=}0.10$ & 68.5 & 46.4 & 88.6 & 36.6 & 80.7 \\
\bottomrule
\end{tabular}
\vspace{-10pt}
\end{table}

Table~\ref{tab:objectives_lambda} compares the training objectives for the multilingual model. The hybrid objective is markedly better on both the retrieval tasks, while TT-MSE performance is better for ZSC. Distilling from the audio encoder alone (AT-MSE, AT-NCE) is clearly worse for all tasks compared to distillation from the text encoders (TT-MSE and Hybrid loss). Though we did not report per-language performance metrics, we note that the trends hold across all languages.
Table~\ref{tab:objectives_lambda} also reports performance as a
function of $\lambda$ for the hybrid objective. At $\lambda{=}0$ the model aligns only with the text encoder, but slightly different from TT-MSE as our hybrid objective uses CS instead of MSE for the text encoder, and at $\lambda{=}1$ the objective reduces to AT-NCE, which performs poorly. Intermediate values improve retrieval substantially at the expense of ZSC. 
We hypothesize that a small non-zero $\lambda$ minimizes the \emph{modality gap} (discussed in Section~\ref{sec:modality_gap_analysis}) between the audio and text encoder spaces leading to a better retrieval performance.

\noindent Table~\ref{tab:objectives_lambda} also shows the performance of GLAP \cite{dinkel2025glap} on the same test set. Results show a strong zero-shot performance across Indic languages in the retrieval tasks. In particular, the model is better than the best Indic-CLAP model in both the retrieval tasks, but the ZSC performance is poor. 
Finetuning the text encoder of GLAP model with the proposed objectives and using translated Indic language captions data is found to improve all performance metrics. The ZSC performance of GLAP fine-tuned using TT-MSE is closer to that of Indic-CLAP and the performance in retrieval tasks approaches that of GLAP teacher. Results show that the proposed cross-lingual teacher learning strategy is effective for extending the multilingual capabilities of constrastively learned representation models.


\begin{table}[t]
\centering
\setlength{\tabcolsep}{3pt}
\caption{Performance of the multilingual models on UrbanSound8K, and Clotho datasets, averaged over the nine languages.}
\label{tab:generalisation}
\vspace{3pt}
\begin{tabular}{l c cc cc}
\toprule
 & \textbf{US8K} & \multicolumn{2}{c}{\textbf{Clotho A$\rightarrow$T}}
 & \multicolumn{2}{c}{\textbf{Clotho T$\rightarrow$A}} \\
\cmidrule(lr){3-4}\cmidrule(lr){5-6}
\textbf{Objective} & \textbf{ZSC} & \textbf{R@1} & \textbf{R@10}
                   & \textbf{R@1} & \textbf{R@10} \\
\midrule
TT-MSE & 73.0 & 17.8 & 58.2 & 14.5 & 53.0 \\
AT-MSE & 65.1 &  8.3 & 38.0 &  6.1 & 30.7 \\
AT-NCE & 54.4 & 10.1 & 41.8 & 12.0 & 44.0 \\
\rowcolor{blue!8}
\textbf{Hybrid $\lambda{=}0.10$} & \textbf{76.3} & \textbf{17.9} & \textbf{58.3}
 & \textbf{15.0} & \textbf{53.4} \\
\midrule
\rowcolor{gray!12}
CLAP (teacher) & 79.2 & 29.6 & 74.9 & 22.9 & 68.3 \\
\bottomrule
\end{tabular}
\vspace{-10pt}
\end{table}

\noindent Table~\ref{tab:generalisation} evaluates the multilingual models on
UrbanSound8K \cite{salamon2014us8k} 
and Clotho \cite{drossos2020clotho}. 
On both benchmarks, the hybrid objective perfors better than the rest, although the performance difference to TT-MSE is $\leq 0.5\%$ on retrieval tasks. The audio teacher objectives perform poorer than TT-MSE and the Hybrid objective. There exists a performance gap with respect to the teacher model in retrieval tasks using Clotho, unlike the AudioCaps results in Table~\ref{tab:objectives_lambda}. We attribute this to the generalization gap. CLAP was trained on pooled AudioCaps and Clotho captions, while the Indic-CLAP models were trained only on AudioCaps.

\vspace{-3pt}
\subsection{Modality Gap Analysis}
\vspace{-2pt}
\label{sec:modality_gap_analysis}

\begin{table}[t]
\centering
\small
\setlength{\tabcolsep}{3pt}
\caption{Modality gap, measured for different models, studied against alignment to the text and audio encoders of CLAP. 
}
\label{tab:gap_measured}
\vspace{5pt}
\begin{tabular}{lccccc}
\toprule
 & \multicolumn{2}{c}{\textbf{Centroids}}
 & \multicolumn{2}{c}{\textbf{Matched pairs}} & \\
\cmidrule(lr){2-3}\cmidrule(lr){4-5}
\textbf{Space} & $\rightarrow$txt & $\rightarrow$aud & $\rightarrow$txt & $\rightarrow$aud & \textbf{Gap} \\
\midrule
CLAP-text  & $1.00$  & $-0.34$ & $1.00$ & $0.22$ & $1.00$ \\
CLAP-audio & $-0.34$ & $1.00$  & $0.22$ & $1.00$ & $0.00$ \\
\midrule
TT-MSE & $1.00$  & $-0.35$ & $0.91$ & $0.22$ & $+1.02$ \\
AT-MSE & $-0.34$ & $1.00$  & $0.25$ & $0.82$ & $-0.07$ \\
AT-NCE & $0.04$ & $0.52$ & $0.47$ & $0.57$ & $+0.68$ \\
\rowcolor{blue!8}
Hybrid & $0.99$  & $-0.29$ & $0.88$ & $\mathbf{0.27}$ & $+0.96$ \\
\bottomrule
\end{tabular}
\vspace{-10pt}
\end{table}

Multimodal representation spaces learned contrastively often exhibit a misalignment between modalities, referred to as the \emph{modality gap} \cite{pmlr-v280-yaras25a, liang2022mind}. We quantify it by the CS between the
modality centroids (pooled over all languages) and the mean CS over matched pairs. The \textit{Gap} is measured as the projection on to the audio$\rightarrow$text centroid axis with value $0.0$ and $1.0$ representing the audio and text poles, respectively. 
Results in Table~\ref{tab:gap_measured} show that the centroids of CLAP's text and audio encoders
lie in opposite hemispheres (negative centroid CS) with low matched-pair CS. The TT-MSE student aligns
well with the CLAP text encoder (Gap $=+1.02$), inheriting the poor alignment with the audio encoder (centroid CS $=-0.35$), giving strong classification but poor retrieval performance. AT-MSE is the mirror image, aligning with the audio encoder (Gap $=-0.07$) and losing text-side concept structure, degrading both the tasks. Note that though the centroids align, matched pair CS is poor reflecting the difficulty of distillation between the text and audio encoders using MSE. Hybrid objective marginally improves the centroid and matched pair alignment with the audio encoder of CLAP. We attribute the better performance on retrieval tasks, reported in Tables~\ref{tab:objectives_lambda} and \ref{tab:generalisation} to this improved cross-modal alignment. 

\vspace{-3pt}
\section{Conclusion}
\vspace{-2pt}
We presented Indic-CLAP, the first text--audio model trained and evaluated for Indic languages, extending CLAP-based audio understanding to nine languages using machine-translated captions. Only a multilingual text encoder was trained using distillation, keeping CLAP's audio encoder frozen. Four training objectives were investigated, in which a hybrid distillation using both the encoders of CLAP is found to be effective, often exceeding its teacher's performance on retrieval tasks for in-domain audio. Our analysis shows that the improvements can be attributed to the reduced \textit{modality gap}. Translation quality is a key challenge and we find that LLM-based approaches clearly outperform the traditional NMT approaches.



\newpage
\section*{Compliance with Ethical Standards}
This work used only publicly available datasets (AudioCaps, ESC-50,
UrbanSound8K, Clotho) under their respective licences. Label verification was performed by
volunteer native speakers who gave informed consent; no personal data was
collected and no experiments with human or animal subjects were conducted.

\balance
\footnotesize
\bibliographystyle{IEEEbib}
\bibliography{refs}

@inproceedings{elizalde2023clap,
    author    = {Elizalde, Benjamin and Deshmukh, Soham and
                 Al Ismail, Mahmoud and Wang, Huaming},
    title     = {{CLAP}: Learning Audio Concepts from Natural Language
                 Supervision},
    booktitle = {Proceedings of the IEEE International Conference on
                 Acoustics, Speech and Signal Processing (ICASSP)},
    pages     = {1--5},
    year      = {2023},
    doi       = {10.1109/ICASSP49357.2023.10095889}
}

@inproceedings{wu2023largescaleclap,
    author    = {Wu, Yusong and Chen, Ke and Zhang, Tianyu and Hui, Yuchen
                 and Berg-Kirkpatrick, Taylor and Dubnov, Shlomo},
    title     = {Large-Scale Contrastive Language-Audio Pretraining with
                 Feature Fusion and Keyword-to-Caption Augmentation},
    booktitle = {Proceedings of the IEEE International Conference on
                 Acoustics, Speech and Signal Processing (ICASSP)},
    pages     = {1--5},
    year      = {2023},
    doi       = {10.1109/ICASSP49357.2023.10095969}
}

@inproceedings{deshmukh2023pengi,
    author    = {Deshmukh, Soham and Elizalde, Benjamin and Singh, Rita
                 and Wang, Huaming},
    title     = {Pengi: An Audio Language Model for Audio Tasks},
    booktitle = {Advances in Neural Information Processing Systems
                 (NeurIPS)},
    volume    = {36},
    year      = {2023}
}

@article{dinkel2025glap,
    author  = {Dinkel, Heinrich and Yan, Zhiyong and Wang, Tianzi and
               Wang, Yongqing and Sun, Xingwei and Niu, Yadong and
               Liu, Jizhong and Li, Gang and Zhang, Junbo and Luan, Jian},
    title   = {{GLAP}: General Contrastive Audio-Text Pretraining across
               Domains and Languages},
    journal = {arXiv preprint arXiv:2506.11350},
    year    = {2025}
}

@inproceedings{radford2021clip,
    author    = {Radford, Alec and Kim, Jong Wook and Hallacy, Chris and
                 Ramesh, Aditya and Goh, Gabriel and Agarwal, Sandhini and
                 Sastry, Girish and Askell, Amanda and Mishkin, Pamela and
                 Clark, Jack and Krueger, Gretchen and Sutskever, Ilya},
    title     = {Learning Transferable Visual Models from Natural Language
                 Supervision},
    booktitle = {Proceedings of the 38th International Conference on
                 Machine Learning (ICML)},
    pages     = {8748--8763},
    year      = {2021},
    url       = {https://proceedings.mlr.press/v139/radford21a.html}
}

@inproceedings{carlsson2022mclip,
    author    = {Carlsson, Fredrik and Eisen, Philipp and Rekathati, Faton
                 and Sahlgren, Magnus},
    title     = {Cross-lingual and Multilingual {CLIP}},
    booktitle = {Proceedings of the Thirteenth Language Resources and
                 Evaluation Conference (LREC)},
    pages     = {6848--6854},
    year      = {2022},
    url       = {https://aclanthology.org/2022.lrec-1.739}
}

@inproceedings{conneau2020xlmr,
    author    = {Conneau, Alexis and Khandelwal, Kartikay and Goyal,
                 Naman and Chaudhary, Vishrav and Wenzek, Guillaume and
                 Guzm{\'a}n, Francisco and Grave, {\'E}douard and Ott,
                 Myle and Zettlemoyer, Luke and Stoyanov, Veselin},
    title     = {Unsupervised Cross-lingual Representation Learning at
                 Scale},
    booktitle = {Proceedings of the 58th Annual Meeting of the Association
                 for Computational Linguistics (ACL)},
    pages     = {8440--8451},
    year      = {2020},
    doi       = {10.18653/v1/2020.acl-main.747}
}

@article{khanuja2021muril,
    author  = {Khanuja, Simran and Kakwani, Divyanshu and Golla, Satwick
               and {N}, Gokul and Bhatt, Amir and Paranjape, Manish and
               Dave, Shruti and Kumar, Saurabh and Dabre, Raj and
               Kunchukuttan, Anoop and others},
    title   = {{MuRIL}: Multilingual Representations for Indian Languages},
    journal = {arXiv preprint arXiv:2103.10730},
    year    = {2021}
}

@inproceedings{kakwani2020indicnlpsuite,
    author    = {Kakwani, Divyanshu and Kunchukuttan, Anoop and Golla,
                 Satwick and {N. C.}, Gokul and Bhatt, Amir and Khapra,
                 Mitesh M. and Kumar, Pratyush},
    title     = {{IndicNLPSuite}: Monolingual Corpora, Evaluation
                 Benchmarks and Pre-trained Multilingual Language Models
                 for Indian Languages},
    booktitle = {Findings of the Association for Computational Linguistics:
                 EMNLP 2020},
    pages     = {4948--4961},
    year      = {2020},
    doi       = {10.18653/v1/2020.findings-emnlp.445}
}

@inproceedings{kim2019audiocaps,
    author    = {Kim, Chris Dongjoo and Kim, Byeong and Lee, Hyunmin and
                 Kim, Gunhee},
    title     = {{AudioCaps}: Generating Captions for Audios in the Wild},
    booktitle = {Proceedings of the 2019 Conference of the North American
                 Chapter of the Association for Computational Linguistics:
                 Human Language Technologies (NAACL-HLT)},
    pages     = {119--132},
    year      = {2019},
    doi       = {10.18653/v1/N19-1011}
}

@inproceedings{piczak2015esc,
    author    = {Piczak, Karol J.},
    title     = {{ESC}: Dataset for Environmental Sound Classification},
    booktitle = {Proceedings of the 23rd ACM International Conference on
                 Multimedia (ACM MM)},
    pages     = {1015--1018},
    year      = {2015},
    doi       = {10.1145/2733373.2806390}
}

@inproceedings{gemmeke2017audioset,
    author    = {Gemmeke, Jort F. and Ellis, Daniel P. W. and Freedman,
                 Dylan and Jansen, Aren and Lawrence, Wade and Moore, R.
                 Channing and Plakal, Manoj and Ritter, Marvin},
    title     = {{AudioSet}: An Ontology and Human-Labeled Dataset for
                 Audio Events},
    booktitle = {Proceedings of the IEEE International Conference on
                 Acoustics, Speech and Signal Processing (ICASSP)},
    pages     = {776--780},
    year      = {2017},
    doi       = {10.1109/ICASSP.2017.7952261}
}

@inproceedings{tiedemann2020opus,
    author    = {Tiedemann, J{\"o}rg and Thottingal, Santhosh},
    title     = {{OPUS-MT} --- Building Open Translation Services for the
                 World},
    booktitle = {Proceedings of the 22nd Annual Conference of the European
                 Association for Machine Translation (EAMT)},
    pages     = {479--480},
    year      = {2020}
}

@article{openai2024gpt4,
    author  = {{OpenAI}},
    title   = {{GPT-4} Technical Report},
    journal = {arXiv preprint arXiv:2303.08774},
    year    = {2024}
}

@inproceedings{liang2022mind,
    author    = {Liang, Victor Weixin and Zhang, Yuhui and Kwon, Yongchan
                 and Yeung, Serena and Zou, James},
    title     = {Mind the Gap: Understanding the Modality Gap in
                 Multi-modal Contrastive Representation Learning},
    booktitle = {Advances in Neural Information Processing Systems
                 (NeurIPS)},
    volume    = {35},
    pages     = {17612--17625},
    year      = {2022}
}

@misc{sun2025raclap,
      title={RA-CLAP: Relation-Augmented Emotional Speaking Style Contrastive Language-Audio Pretraining For Speech Retrieval}, 
      author={Haoqin Sun and Jingguang Tian and Jiaming Zhou and Hui Wang and Jiabei He and Shiwan Zhao and Xiangyu Kong and Desheng Hu and Xinkang Xu and Xinhui Hu and Yong Qin},
      year={2025},
      eprint={2505.19437},
      archivePrefix={arXiv},
      primaryClass={cs.SD},
      url={https://arxiv.org/abs/2505.19437}, 
}

@inproceedings{sun25c_interspeech,
  title     = {{RA-CLAP: Relation-Augmented Emotional Speaking Style Contrastive Language-Audio Pretraining For Speech Retrieval}},
  author    = {Haoqin Sun and Jingguang Tian and Jiaming Zhou and Hui Wang and Jiabei He and Shiwan Zhao and Xiangyu Kong and Desheng Hu and Xinkang Xu and Xinhui Hu and Yong Qin},
  year      = {2025},
  booktitle = {{Interspeech 2025}},
  pages     = {2995--2999},
  doi       = {10.21437/Interspeech.2025-548},
  issn      = {2958-1796},
}

@inproceedings{yan24_interspeech,
  title     = {{Bridging Language Gaps in Audio-Text Retrieval}},
  author    = {Zhiyong Yan and Heinrich Dinkel and Yongqing Wang and Jizhong Liu and Junbo Zhang and Yujun Wang and Bin Wang},
  year      = {2024},
  booktitle = {{Interspeech 2024}},
  pages     = {1675--1679},
  doi       = {10.21437/Interspeech.2024-420},
  issn      = {2958-1796},
}

@misc{cousin2023multilingualaudiocaptioningusing,
      title={Multilingual Audio Captioning using machine translated data}, 
      author={Matéo Cousin and Étienne Labbé and Thomas Pellegrini},
      year={2023},
      eprint={2309.07615},
      archivePrefix={arXiv},
      primaryClass={cs.SD},
      url={https://arxiv.org/abs/2309.07615}, 
}

@article{liu2019roberta,
  author  = {Liu, Yinhan and Ott, Myle and Goyal, Naman and Du, Jingfei
             and Joshi, Mandar and Chen, Danqi and Levy, Omer and
             Lewis, Mike and Zettlemoyer, Luke and Stoyanov, Veselin},
  title   = {{RoBERTa}: A Robustly Optimized {BERT} Pretraining Approach},
  journal = {arXiv preprint arXiv:1907.11692},
  year    = {2019}
}

@article{koepke2022retrieval,
  author  = {Koepke, A. Sophia and Oncescu, Andreea-Maria and
             Henriques, Jo{\~a}o F. and Akata, Zeynep and Albanie, Samuel},
  title   = {Audio Retrieval with Natural Language Queries:
             {A} Benchmark Study},
  journal = {IEEE Transactions on Multimedia},
  year    = {2022},
  doi     = {10.1109/TMM.2022.3149712}
}

@inproceedings{liu2023audioldm,
  author    = {Liu, Haohe and Chen, Zehua and Yuan, Yi and Mei, Xinhao
               and Liu, Xubo and Mandic, Danilo and Wang, Wenwu and
               Plumbley, Mark D.},
  title     = {{AudioLDM}: Text-to-Audio Generation with Latent
               Diffusion Models},
  booktitle = {Proceedings of the International Conference on
               Machine Learning (ICML)},
  year      = {2023}
}

@inproceedings{ghosal2023tango,
  author    = {Ghosal, Deepanway and Majumder, Navonil and Mehrish, Ambuj
               and Poria, Soujanya},
  title     = {Text-to-Audio Generation using Instruction-Tuned {LLM}
               and Latent Diffusion Model},
  booktitle = {Proceedings of the 31st ACM International Conference
               on Multimedia (MM)},
  year      = {2023}
}

@article{oord2018cpc,
    author  = {van den Oord, A{\"a}ron and Li, Yazhe and Vinyals, Oriol},
    title   = {Representation Learning with Contrastive Predictive Coding},
    journal = {arXiv preprint arXiv:1807.03748},
    year    = {2018}
}

@inproceedings{chen2022htsat,
    author    = {Chen, Ke and Du, Xingjian and Zhu, Bilei and Ma, Zejun
                 and Berg-Kirkpatrick, Taylor and Dubnov, Shlomo},
    title     = {{HTS-AT}: A Hierarchical Token-Semantic Audio Transformer
                 for Sound Classification and Detection},
    booktitle = {Proceedings of the IEEE International Conference on
                 Acoustics, Speech and Signal Processing (ICASSP)},
    pages     = {646--650},
    year      = {2022},
    doi       = {10.1109/ICASSP43922.2022.9746312}
}

@InProceedings{pmlr-v280-yaras25a,
  title = 	 {Explaining and Mitigating the Modality Gap in Contrastive Multimodal Learning},
  author =       {Yaras, Can and Chen, Siyi and Wang, Peng and Qu, Qing},
  booktitle = 	 {Conference on Parsimony and Learning},
  pages = 	 {1365--1387},
  year = 	 {2025},
  editor = 	 {Chen, Beidi and Liu, Shijia and Pilanci, Mert and Su, Weijie and Sulam, Jeremias and Wang, Yuxiang and Zhu, Zhihui},
  volume = 	 {280},
  series = 	 {Proceedings of Machine Learning Research},
  month = 	 {24--27 Mar},
  publisher =    {PMLR},
  url = 	 {https://proceedings.mlr.press/v280/yaras25a.html},

}

@inproceedings{salamon2014us8k,
  title     = {A Dataset and Taxonomy for Urban Sound Research},
  author    = {Salamon, Justin and Jacoby, Christopher and Bello, Juan Pablo},
  booktitle = {Proc. ACM Int. Conf. Multimedia},
  pages     = {1041--1044},
  year      = {2014}
}

@inproceedings{drossos2020clotho,
  title     = {Clotho: an Audio Captioning Dataset},
  author    = {Drossos, Konstantinos and Lipping, Samuel and Virtanen, Tuomas},
  booktitle = {Proc. IEEE ICASSP},
  pages     = {736--740},
  year      = {2020}
}

@article{duquenne2023sonar,
    author  = {Duquenne, Paul-Ambroise and Schwenk, Holger and Sagot, Beno{\^i}t},
    title   = {{SONAR}: Sentence-Level Multimodal and Language-Agnostic
               Representations},
    journal = {arXiv preprint arXiv:2308.11466},
    year    = {2023}
}

\end{document}